*Article*

# Predicting Neoadjuvant Chemotherapy Response in Triple-Negative Breast Cancer Using Pre-Treatment Histopathologic Images

**Hikmat Khan [1,*], Ziyu Su [1], Huina Zhang [2], Yihong Wang [3], Bohan Ning [4], Shi Wei [4], Hua Guo [4], Zaibo Li [1] and Muhammad Khalid Khan Niazi [1]**

[1] Department of Pathology, College of Medicine, Wexner Medical Center, The Ohio State University, Columbus, OH, 43210, USA; ziyu.su@osumc.edu (Z.S.); zaibo.li@osumc.edu (Z.L.); khalid.niazi@osumc.edu (M.K.K.N.)
[2] Department of Pathology, University of Rochester Medical Center, Rochester, NY 14642, USA; huina_zhang@urmc.rochester.edu
[3] Department of Pathology and Laboratory Medicine, Warren Alpert Medical School of Brown University, Lifespan Academic Medical Center, Providence, RI 02903, USA; yihong_wang@brown.edu
[4] Department of Pathology, University of Alabama at Birmingham, Birmingham, AL, USA; bning@uab.edu (B.N.); swei@uabmc.edu (S.W.); huaguo@uabmc.edu (H.G.)
* Correspondence: hikmat.khan@osumc.edu

## Simple Summary

Triple-negative breast cancer (TNBC) is a fast-growing and hard-to-treat form of breast cancer that does not respond to hormone therapies. Although chemotherapy before surgery—called neoadjuvant chemotherapy (NACT)—is the standard treatment, not all patients benefit from it. In this study, we developed an artificial intelligence (AI) model that analyzes routine biopsy slides to predict which patients are likely to respond well to NACT. The model showed strong performance in both internal and external patient groups and focused on tumor regions rich in immune cells, such as CD8+ T, CD163+ and PD-L1. This approach could help personalize treatment, reduce unnecessary side effects, and guide more effective care for patients with TNBC.

## Abstract

Triple-negative breast cancer (TNBC) remains a major clinical challenge due to its aggressive behavior and lack of targeted therapies. Accurate early prediction of response to neoadjuvant chemotherapy (NACT) is essential for guiding personalized treatment strategies and improving patient outcomes. In this study, we present an attention-based multiple instance learning (MIL) framework designed to predict pathologic complete response (pCR) directly from pre-treatment hematoxylin and eosin (H&E)-stained biopsy slides. The model was trained on a retrospective in-house cohort of 174 TNBC patients and externally validated on an independent cohort (n = 30). It achieved a mean area under the curve (AUC) of 0.85 during five-fold cross-validation and 0.78 on external testing, demonstrating robust predictive performance and generalizability. To enhance model interpretability, attention maps were spatially co-registered with multiplex immunohistochemistry (mIHC) data stained for PD-L1, CD8+ T cells, and CD163+ macrophages. The attention regions exhibited moderate spatial overlap with immune-enriched areas, with mean Intersection over Union (IoU) scores of 0.47 for PD-L1, 0.45 for CD8+ T cells, and 0.46 for CD163+ macrophages. The presence of these biomarkers in high-attention regions supports their biological relevance to NACT response in TNBC. This not only improves model interpretability but may also inform future efforts to identify clinically actionable histological biomarkers directly from H&E-stained biopsy slides, further supporting the utility of this approach for accurate NACT response prediction and advancing precision oncology in TNBC.

## 1. Introduction

TNBC accounts for approximately 15–20% of all invasive breast cancers worldwide, corresponding to an estimated 200,000–300,000 new cases annually [1]. It is characterized by the absence of estrogen receptor (ER), progesterone receptor (PR), and human epidermal growth factor receptor 2 (HER2) expression or gene amplification [2-7]. This lack of targetable receptors limits systemic treatment options, making neoadjuvant chemotherapy (NACT) the standard initial approach for early-stage disease [8,9]. The goal of NACT is to achieve pathologic complete response (pCR; ypT0N0, indicating no residual invasive carcinoma in the breast or lymph nodes) [10,11] and to downstage tumors, thereby improving surgical outcomes and enabling breast-conserving surgery in patients who might otherwise require mastectomy [12].

Approximately 40 to 50% of patients achieve a pCR [4,13-17], a critical surrogate endpoint that is strongly associated with improved survival [18]. In contrast, patients with residual disease (i.e., non-pCR, who did not achieve pCR) face higher rates of relapse and worse overall survival [8,19]. This underscores the urgent need for early prediction of NACT response to guide clinical decisions, including tailoring treatments, optimizing surgical planning, and avoiding unnecessary toxicity from ineffective regimens [20-22]. Furthermore, early identification of patients that will likely be non-pCR could also support timely consideration of alternative therapies or clinical trial enrollment [18,23]. However, due to TNBC's aggressive biology and the lack of reliable predictive biomarkers in clinical practice, outcome prediction remains a major unmet need [17,20,24-27]. In this study, we present an attention-based multiple instance learning (MIL) framework to predict NACT response in TNBC patients from pre-treatment H&E-stained biopsy images. The main contributions of our work are as follows:

1. We employed an attention-based MIL framework that utilizes pre-treatment H&E-stained images to predict response (i.e., either pCR or non-pCR) to NACT in TNBC patients. Our framework demonstrates strong average predictive performance on an in-house cohort of 174 TNBC patients—an accuracy of 82%, AUC of 0.86, F1-score of 0.84, sensitivity of 0.85, specificity of 0.81, and precision of 0.80 based on five-fold cross-validation—outperforming a traditional model that relies only on clinical data.
2. We evaluated our attention-based MIL framework on an independent cohort of 30 TNBC patients (12 pCR and 18 non-pCR), achieving an accuracy of 76%, AUC of 0.78, F1-score of 0.67, sensitivity of 0.72, specificity of 0.73, and precision of 0.81, demonstrating its generalizability and potential for clinical utility.
3. To quantitatively evaluate the biological plausibility of the model's attention, we computed the IoU between our model's attention regions in H&E-stained biopsy slides and corresponding regions in co-registered multiplex mIHC data stained for PD-L1, $CD8^{+}$ T cells, and $CD163^{+}$ macrophages. Notably, we found that the model attention regions showed moderate overlap with these biomarkers, with IoU scores of 0.47 for PD-L1, 0.45 for $CD8^{+}$ T cells, and 0.46 for $CD163^{+}$ macrophages. The presence of these biomarkers in high-attention regions highlights their biological relevance to NACT response in TNBC and may improve model interpretability while informing future efforts to identify clinically actionable histological biomarkers directly from H&E-stained images.

In summary, these contributions underscore the potential of the proposed attention-based MIL framework applied to pre-treatment H&E-stained biopsy slides for predicting NACT response in TNBC. The proposed framework demonstrates strong predictive performance, biological interpretability via immune biomarker alignment, and generalizability across independent cohorts, highlighting its translational relevance for clinical decision support in precision oncology.

## 2. Related Work

Many studies have investigated whether standard clinicopathological features—such as tumor size, histologic grade, molecular subtype, and lymph node involvement—can predict response to NACT in TNBC. However, findings remain inconsistent [28]. While some reports suggest that smaller, lower-grade, and node-negative tumors are more likely to achieve a pCR, others find no significant association, indicating that these conventional features alone lack sufficient predictive power [29-36].

In the absence of reliable clinical biomarkers, machine learning and artificial intelligence (AI) approaches have emerged as promising tools for predicting NACT outcomes. Recent studies have focused heavily on radiomics and deep learning methods applied to medical imaging, particularly magnetic resonance imaging (MRI) and ultrasound. For example, Zhou et al. [37] employed a deep learning model on multiparametric MRI (DCE-MRI and DWI) and achieved an AUC of 0.86, suggesting early-treatment-phase imaging may capture predictive signatures. In contrast, Golden et al. [38] reported a more modest AUC of 0.68, potentially due to smaller sample size (n = 60) or suboptimal feature selection. Jiang et al. [39] used ultrasound-based radiomics in a cohort of 592 TNBC patients, achieving an AUC of 0.93 and an accuracy of 0.84.

Several studies combined imaging with clinical variables to improve accuracy. For instance, Xu et al. [40] integrated MRI with clinicopathological data (AUC = 0.76), while Jimenez et al. [41] incorporated tumor-infiltrating lymphocytes (AUC = 0.71). These approaches suggest that multimodal data integration may better capture tumor heterogeneity, although challenges in interpretability persist.

Despite their potential, imaging-based models depend on modality access, protocol consistency, and cost-intensive workflows, limiting their scalability in routine clinical settings. In contrast, H&E-stained biopsy images are universally available and standardized. Yet, histopathology-based deep learning models remain relatively underexplored. Recently, Savitri et al. [42] pioneered deep learning on H&E-stained slides (AUC = 0.75), offering a cost-effective alternative to imaging. Huang et al. [43] proposed IMPRESS, an AI pipeline integrating H&E with mIHC markers (PD-L1, CD8+, CD163+), reporting an AUC of 0.8975 for HER2+ and 0.7674 for TNBC, demonstrating that AI-based methods can outperform manual pathologist assessments in predicting NACT response. Hussain et al. [44] explore deep learning advancements in biomarker discovery and multi-omics integration to enhance TNBC management, while highlighting challenges such as model interpretability and limited data availability, and emphasize the importance of multidisciplinary collaboration and continued research.

In summary, while prior studies have leveraged imaging and clinical data to forecast NACT response in TNBC, histopathology-driven AI models offer a cost-effective, scalable, and biologically interpretable alternative. Our study builds on these early efforts by applying a multiple instance learning framework to pre-treatment H&E-stained biopsy slides, enhanced through alignment with immune markers derived from mIHC.

# 3. Materials

This section describes the two cohorts (i.e., datasets) used in this study, including the in-house cohort and the independent validation cohort. Both cohorts are available from the corresponding author upon reasonable request for research purposes. The study was approved by the Institutional Review Board (IRB protocol #2016C0025).

### 3.1. In-House Cohort

In this retrospective study, we included 174 female patients diagnosed with TNBC and treated with NACT at The Ohio State University Wexner Medical Center (OSUWMC) between 2013 and 2020. All patients had documented treatment outcomes, pre-treatment H&E-stained biopsy slides with tumor regions greater than 0.1 cm, and corresponding clinical data. Among these patients, 81 achieved a pCR, while 93 were categorized as non-pCR. Additionally, a subset of 64 patients had pre-NACT mIHC slides stained for CD8+ T cells, CD163+ macrophages, and PD-L1 biomarkers. This subset enabled downstream biomarker analysis and interpretability validation. To ensure robust model evaluation, we performed five-fold cross-validation at the patient level using stratified sampling, maintaining balanced pCR and non-pCR ratios across the training, validation, and test sets. The training set was used for model development, the validation set guided hyperparameter tuning and early stopping, and the test set was used for fold-level performance evaluation.

### 3.2. Independent Validation Cohort

To assess model generalizability, we included an independent cohort of 30 TNBC patients (12 achieved pCR; 18 were non-pCR) from The University of Texas MD Anderson Cancer Center collected between 2017 and 2022. Each patient had pre-treatment H&E-stained biopsy slides, and this cohort was used exclusively for external testing. Table 1 summarizes the distribution of pCR and non-pCR cases across both cohorts (see Supplementary Table S1 for additional details on individual cohort data).

**Table 1.** Distribution of patient cohorts used in this study, showing the number of cases with pathologic complete response (pCR) and pathologic incomplete response (non-pCR) for both the in-house Ohio State University Wexner Medical Center cohort and the independent MD Anderson cancer center cohort.

| Cohorts | pCR Cases | Non-pCR Cases | Total |
|---|---|---|---|
| OSU-Wexner Medical Center (in-house) | 81 | 93 | 174 |
| MD Anderson Cancer Center (independent) | 12 | 18 | 30 |

# 4. Method

We employed an attention-based MIL framework to predict response to NACT, distinguishing between pCR and non-pCR outcomes using pre-treatment H&E-stained biopsy slides [45]. Figure 1 illustrates an overview of the framework, which consists of four main stages: (1) tissue patch extraction, (2) patch-level feature encoding, (3) attention-based aggregation, and (4) slide-level classification. We begin with a brief overview of the attention-based weakly supervised learning strategy, followed by a detailed description of each stage in the subsequent subsections.

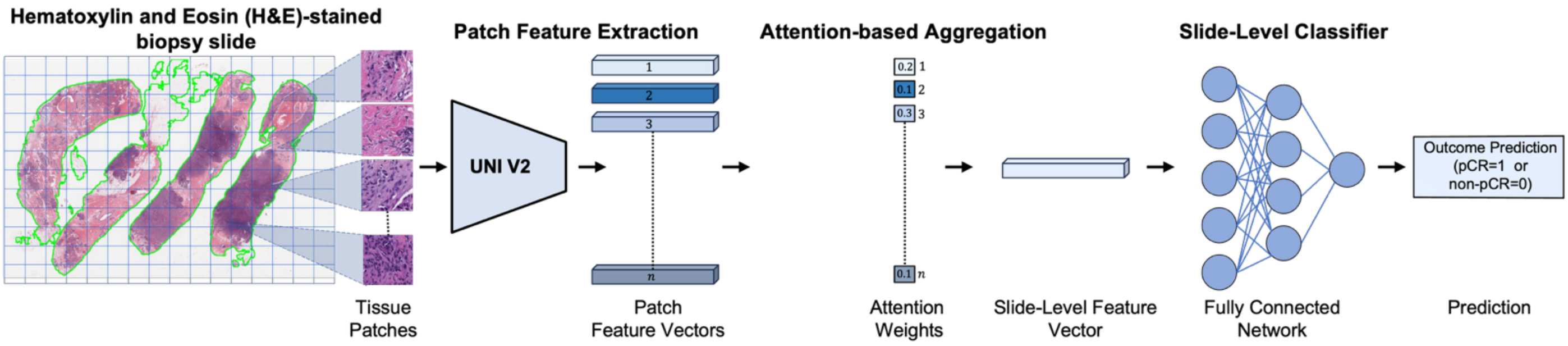

**Figure 1.** Overview of the pipeline for predicting pathologic complete response (pCR) to neoadjuvant chemotherapy (NACT) in triple-negative breast cancer (TNBC) using pre-treatment H&E-stained biopsy slides. First, the H&E-stained slide is segmented and divided into a grid to extract tissue patches. Each patch is then encoded into a feature vector using a pretrained deep learning encoder (i.e., UNI v2 [46], a general-purpose, self-supervised pathology foundation model trained on 1.2 million histopathology slides). These patch-level features are aggregated via an attention mechanism [45] that assigns greater weight to the most informative regions, resulting in a slide-level feature representation. A fully connected neural network classifier then utilizes the slide-level feature representation to predict the likelihood of a complete response (pCR) or non-response (non-pCR) to NACT for each patient.

### *4.1. Overview*

Each H&E-stained biopsy slide is divided into non-overlapping patches (also referred to as instances or tiles). Patches from the same biopsy slide are grouped into a single "bag" with only a slide-level label (pCR or non-pCR) assigned. The model is trained to identify and attend to the most informative patches within the bag of each slide that contribute to the overall NACT response prediction.

### *4.2. Patch Extraction and Feature Encoding*

In the MIL framework, each H&E-stained biopsy slide is partitioned into non-overlapping 512×512-pixel patches at 40× magnification (0.25 μm/pixel resolution). Patches from the same biopsy slide are grouped into a single "bag". Each patch in the given bag is then passed through a pretrained UNI v2 (a general-purpose, self-supervised pathology foundation model trained on 1.2 million histopathology slides) [46] to extract discriminative feature $h_i \in \mathbb{R}^d$ for each patch $i$, where d = 1536 denotes the feature dimensionality, corresponding to the output of the penultimate layer of the UNI v2 encoder [46,47]. The resulting bag of patch-level feature vectors serves as the input to the next stage, where attention-based aggregation enables the model to focus on the most informative patches for the overall prediction of pCR to NACT.

### *4.3. Attention-Based Aggregation*

In this stage, an attention mechanism is employed to learn attention weights for each patch feature vector (i.e., $\alpha_i \in [0,1]$, satisfying $\sum \alpha_i = 1$) [45], representing the contribution (or importance) of each patch to the final slide-level prediction. Then, a slide-level feature vector $z$ is computed using attention-weighted aggregation, formally defined below.

$$z = \sum_{k=1}^{N} \alpha_k h_k$$

where $N$ is the number of patches in the given slide, $h_k$ is the feature vector for the $k$-th patch, and $\alpha_k$ is the corresponding attention weight for the $k$-th patch. The attention weights are computed as follows:

$$\alpha_k = \frac{exp(w^T\, tanh(Vh_k^T))}{\sum_{j=1}^{N} exp(w^T\, tanh(Vh_j^T))}$$

where $w$ and $V$ are learnable parameters of the attention-based MIL framework, while $\alpha_k$ represents the normalized attention weight of the k-th patch in the final prediction. The attention mechanism offers two key benefits: (1) it enhances predictive performance by adaptively focusing on the most relevant morphological features, and (2) it provides interpretability through spatial attention maps that highlight histological regions strongly associated with NACT response prediction .

*4.4. Slide-Level Classification*

In this stage, the aggregated slide-level feature vector serves as input to a fully connected network with a final sigmoid activation to estimate the probability of a pCR versus non-pCR response to treatment outcome for NACT in TNBC [48].

*4.5. Class-Weighted Loss Function*

To address the challenge of class imbalance, we employed a class-weighted binary cross-entropy loss function, defined as

$$\mathcal{L} = -\frac{1}{B}\sum_{k=1}^{B}[w_{pCR}.\, y_k\, log\, p_k + w_{non-pCR}.\, (1-y_k)\, log(1-p_k)]$$

where $B$ is the batch size, $y_k \in \{0,1\}$ is the ground-truth label, and $p_k$ is the predicted probability for the $k$-th bag. The class weights $w_{pCR}$ and $w_{non-pCR}$ are assigned to the pCR and non-pCR classes and are computed as follows:

$$w_{pCR} = \frac{N}{2 * N_{pCR}}$$
$$w_{non-pCR} = \frac{N}{2 * N_{non-pCR}}$$

where $N$ is the total sum of $N_{pCR}$ and $N_{non-pCR}$; $N_{pCR}$ is the number of pCR and $N_{non-pCR}$ is the number of non-pCR patients, respectively. The class weighting scheme compensates for the inherent imbalance in NACT response by assigning a higher penalty to misclassification of the minority class, thereby encouraging the model to be more sensitive to underrepresented cases during training.

## 5. Experimental Setup

*5.1. Data Augmentation*

To mitigate overfitting and enhance generalization, patch-level data augmentation was applied during training. Augmentations included random rotations (with ±30°), together with horizontal and vertical flips performed with a probability of 0.5, and color jittering (±0.2 adjustment in brightness, contrast, saturation, and hue) with a probability of 0.25 [49]. These augmentations not only improved robustness to histological and staining variations but also helped reduce the risk of overfitting, particularly in medical imaging tasks with limited sample sizes [50].

*5.2. Training and Implementation Details*

All experiments were implemented in PyTorch v2.7 and executed on an NVIDIA A100 GPU. We used the publicly available Trident library to patchify H&E-stained biopsy slides into non-overlapping 512 × 512-pixel tissue patches at 40× magnification (0.25 μm/pixel resolution). The pretrained UNI-V2 model was used as a feature extractor, producing 1536-dimensional feature embeddings for each patch, which served as inputs to the model. Model training was performed using stochastic gradient descent [51,52] with a learning rate of 0.0001, a weight decay of 0.001, and early stopping (patience of 50 epochs) based on minimum validation loss. Each experiment used a batch size of 1 and trained for up to 1024 epochs. A comprehensive list of fixed hyperparameters is provided in Supplementary Tables S2 and S3, and the hyperparameter search space is detailed in Supplementary Table S4. Optimal values were determined using a grid search strategy.

### *5.3. Baseline Models for Comparison*

To benchmark performance, we compared our model against a set of traditional machine learning classifiers, including logistic regression [53], random forest [54], support vector machines (SVMs) [55], k-nearest neighbors (KNN) [56], and linear discriminant analysis (LDA) [57], trained on clinical data to predict NACT response. The established hyperparameter settings for each method are listed in Supplementary Table S5, while the corresponding search spaces are provided in Supplementary Table S6. Optimal values were selected via the grid search strategy.

### *5.4. Evaluation*

We assessed the model's predictive performance using three primary metrics: accuracy (ACC), area under the receiver operating characteristic curve (AUC-ROC), and F1-score. Accuracy quantifies the proportion of correct predictions, encompassing both true positives (TPs) and true negatives (TNs), out of all cases: it is formally defined as

$$ACC = \frac{TP + TN}{TP + TN + FN + FP}$$

The F1-score provides a balanced measure of model performance by combining precision and recall, making it especially useful in the context of class imbalance: it is formally defined as

$$F1 - score = \frac{2. Precision \,. Recall}{Precision + Recall}$$

Where

$$Precision = \frac{TP}{TP + FP}$$

$$Recall = \frac{TP}{TP + FN}$$

## 6. Results and Discussion

### *6.1. Performance on the In-House Cohort*

We initially evaluated the proposed attention-based MIL framework using five-fold cross-validation on the in-house OSUWMC cohort. As detailed in Table 2, our model achieved a mean accuracy of 0.82, AUC-ROC of 0.86, F1-score of 0.84, sensitivity of 0.85, specificity of 0.81, and precision of 0.80. The corresponding ROC curves and confusion matrices for each test fold are presented in Figures 2 and 3, respectively. An analysis of the confusion matrices indicates a balanced distribution of false positives (predicting pCR when the patient did not achieve it) and false negatives (predicting non-pCR when the patient did achieve pCR). This balanced error profile, combined with consistently high sensitivity and specificity (greater than 0.80), underscores the model's robust discriminative performance and its clinical relevance for reliably predicting the pCR to NACT.

**Table 2.** Performance metrics for predicting pathologic complete response (pCR) versus non-pCR following neoadjuvant chemotherapy (NACT) in triple-negative breast cancer (TNBC) in the in-house cohort, evaluated using five-fold cross-validation. Metrics are reported for each fold (Folds 1–5), along with the mean ± standard deviation across all test folds, demonstrating consistent performance across accuracy, AUC, F1-score, sensitivity, specificity, and precision.

| Folds | Accuracy | AUC | F1-Score | Sensitivity | Specificity | Precision |
|---|---|---|---|---|---|---|
| 1 | 0.83 | 0.88 | 0.88 | 0.89 | 0.82 | 0.78 |
| 2 | 0.78 | 0.83 | 0.78 | 0.78 | 0.78 | 0.78 |
| 3 | 0.83 | 0.91 | 0.86 | 0.90 | 0.80 | 0.75 |
| 4 | 0.83 | 0.85 | 0.88 | 0.89 | 0.82 | 0.78 |
| 5 | 0.83 | 0.81 | 0.80 | 0.78 | 0.84 | 0.89 |
| | 0.82 ± 0.02 | 0.86 ± 0.03 | 0.84 ± 0.04 | 0.85 ± 0.06 | 0.81 ± 0.01 | 0.80 ± 0.05 |

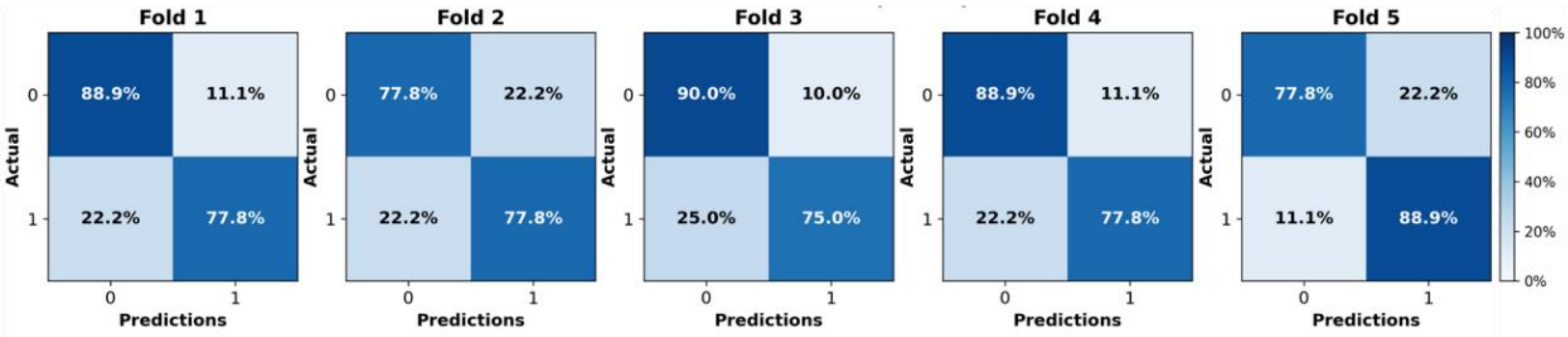

**Figure 2.** Confusion matrices illustrating the model's performance on each test fold of the in-house cohort using five-fold cross-validation.

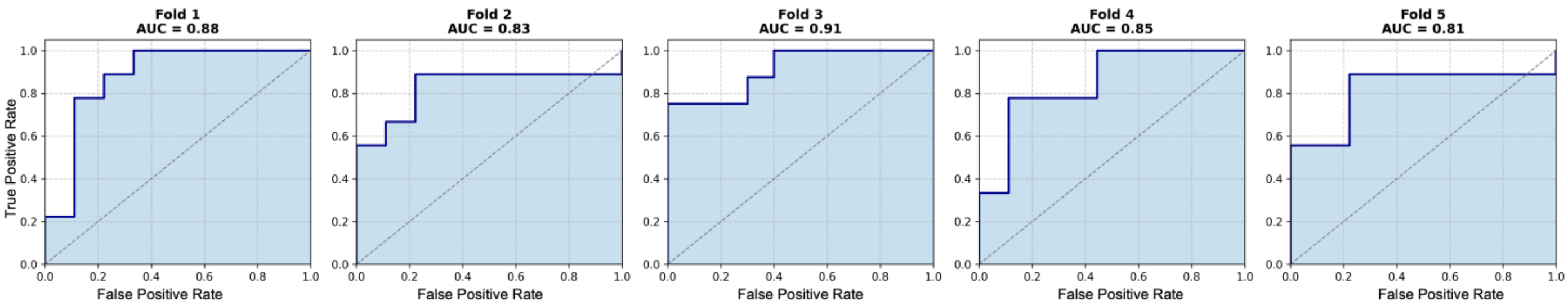

**Figure 3.** Receiver operating characteristic (ROC) curves for each test fold of the in-house cohort using five-fold cross-validation. Area under the curve (AUC) values range from 0.81 to 0.91.

### *6.2. Generalization to External Validation Cohort*

To assess generalizability, the trained models were evaluated on an independent cohort of 30 TNBC cases from the MD Anderson Cancer Center. As shown in Table 3, the model achieved a mean accuracy of 0.76, AUC-ROC of 0.82, and F1-score of 0.77, and a sensitivity of 0.72 and specificity of 0.73 on the independent validation cohort. Although a relative performance drop of approximately 6% was observed, the model retained balanced sensitivity and specificity, indicating strong generalization to out-of-distribution data and supporting its applicability in real-world clinical settings.

**Table 3.** Performance metrics for predicting pathologic complete response (pCR) versus non-pCR following neoadjuvant chemotherapy (NACT) in triple-negative breast cancer (TNBC) are reported for both in-house and independent cohorts. The average metrics are presented as mean ± standard deviation across three independent runs, demonstrating consistent performance in terms of accuracy, AUC, F1-score, sensitivity, specificity, and precision.

| Cohorts | Accuracy | AUC | F1-Score | Sensitivity | Specificity | Precision |
|---|---|---|---|---|---|---|
| OSU-Wexner Medical Center (in-house) | 0.82 ± 0.02 | 0.86 ± 0.03 | 0.84 ± 0.04 | 0.85 ± 0.06 | 0.81 ± 0.01 | 0.80 ± 0.05 |
| MD Anderson Cancer Center (independent) | 0.76 ± 0.03 | 0.78 ± 0.02 | 0.67 ± 0.07 | 0.72 ± 0.11 | 0.73 ± 0.02 | 0.81 ± 0.11 |

### *6.3. Comparison with Classical ML Models*

To contextualize the performance of our framework, we evaluated it against five classical machine learning classifiers trained solely on clinical features: logistic regression [53], random forest [54], SVM [55], KNN [56], and LDA [57]. The clinical features included age, tumor type, HER2 IHC score, HER2 copy number, and HER2 ratio. As shown in Table 4, the best-performing baseline method achieved a maximum AUC of 0.79, which is notably lower than that of our proposed model. This performance gap highlights the advantage of leveraging spatial histopathological features via attention-based MIL rather than depending solely on structured clinical variables for predicting NACT treatment response in TNBC.

**Table 4.** Comparison of the proposed model with classical ML baselines trained on clinical data.

| Model | Accuracy | AUC | F1-Score | Sensitivity | Specificity | Precision |
|---|---|---|---|---|---|---|
| Logistic Regression [53] | 0.57 ± 0.08 | 0.71 ± 0.05 | 0.68 ± 0.05 | 0.89 ± 0.07 | 0.27 ± 0.167 | 0.55 ± 0.07 |
| Random Forest [54] | 0.60 ± 0.05 | 0.72 ± 0.04 | 0.62 ± 0.06 | 0.65 ± 0.08 | 0.56 ± 0.10 | 0.59 ± 0.06 |
| SVM [55] | 0.61 ± 0.05 | 0.67 ± 0.05 | 0.72 ± 0.03 | 1.00 ± 0.00 | 0.22 ± 0.90 | 0.56 ± 0.03 |
| K-Nearest Neighbors [56] | 0.64 ± 0.04 | 0.74 ± 0.04 | 0.70 ± 0.02 | 0.86 ± 0.11 | 0.42 ± 0.17 | 0.20 ± 0.08 |
| Linear Discriminant Analysis [57] | 0.57 ± 0.07 | 0.71 ± 0.04 | 0.67 ± 0.02 | 0.89 ± 0.07 | 0.24 ± 0.18 | 0.54 ± 0.05 |
| Ours | 0.82 ± 0.02 | 0.86 ± 0.03 | 0.84 ± 0.04 | 0.85 ± 0.06 | 0.81 ± 0.01 | 0.80 ± 0.05 |

### *6.4. Attention Map Analysis and Corresponding Biological Insights*

Figures 4 and 5 present attention maps generated by our model for representative test cases corresponding to a patient with pCR and a non-responder (non-pCR), respectively. Visual inspection of the highlighted regions in the H&E-stained slides, co-registered with their corresponding mIHC slides, demonstrated that the model predominantly focused on regions enriched with immune biomarkers—PD-L1 (shown in brown), $CD8^+$ T-cell infiltration (shown in green), and $CD163^+$ macrophages (shown in red). To quantitatively assess the biological plausibility of these attention maps, we computed the Intersection over Union (IoU) between the model-generated attention maps and the spatial distribution of immune biomarkers in the aligned mIHC images. Figure 6 shows the H&E image, the registered mIHC counterpart, the model's attention map, and the masks for $CD8^+$ T cells, $CD163^+$ macrophages, and PD-L1. The immune cell segmentation masks were generated by fine-tuning the CellViT++ model [58] and subsequently used for IoU calculation. Formally, the IoU for PD-L1, for example, is defined as

$$\mathrm{PD-L1\ (IoU)} = \frac{\mathrm{PD-L1\ (mask)} \cap \mathrm{Binarized\ Attention\ Map}}{\mathrm{Total\ number\ of\ PD-L1\ in\ mask}},$$

where $\cap$ denotes the intersection, the numerator represents the overlapping region between the PD-L1 mask and the binarized attention map, and the denominator corresponds to the total area of the PD-L1 mask. We computed the IoU similarly for the other two biomarkers. As summarized in Table 5, the attention maps exhibited moderate spatial overlap with biomarker-enriched regions, with mean IoU scores of 0.47 ± 0.18 for PD-L1, 0.45 ± 0.20 for $CD8^+$ T cells, and 0.46 ± 0.17 for $CD163^+$ macrophages.

**Table 5.** Quantification of PD-L1, $CD8^+$ T, and $CD163^+$ biomarkers using model attention maps. This table presents the mean Intersection over Union (IoU) values between individual cells and the model's attention map. A higher IoU indicates greater presence of the biomarker within the model's attended region.

| Biomarker | IoU (with B-Attention Map) |
|---|---|
| PD-L1 | 0.47 ± 0.18 |
| CD8+ T | 0.45 ± 0.20 |
| CD163+ | 0.46 ± 0.17 |

These IoU results (see Table 5) underscore the biological plausibility of the model's attention mechanism and are consistent with prior studies that highlight the role of immune infiltration and tumor-associated macrophages (TAMs), especially CD163+ M2-polarized subsets, in therapy resistance and poor pCR outcomes [28,59-62]. Notably, even in misclassified samples (Figures 7–9), the model's attention consistently localized to tumor-dense regions, suggesting that its predictions for NACT treatment response are based on morphologically and biologically meaningful patterns.

Given the absence of definitive histological biomarkers for pCR to NACT [34], the biologically grounded interpretability of the attention maps not only enhances transparency but may also informs biomarker discovery directly from H&E images, potentially facilitating reliable and improved NACT response prediction in TNBC patients and advancing precision oncology.

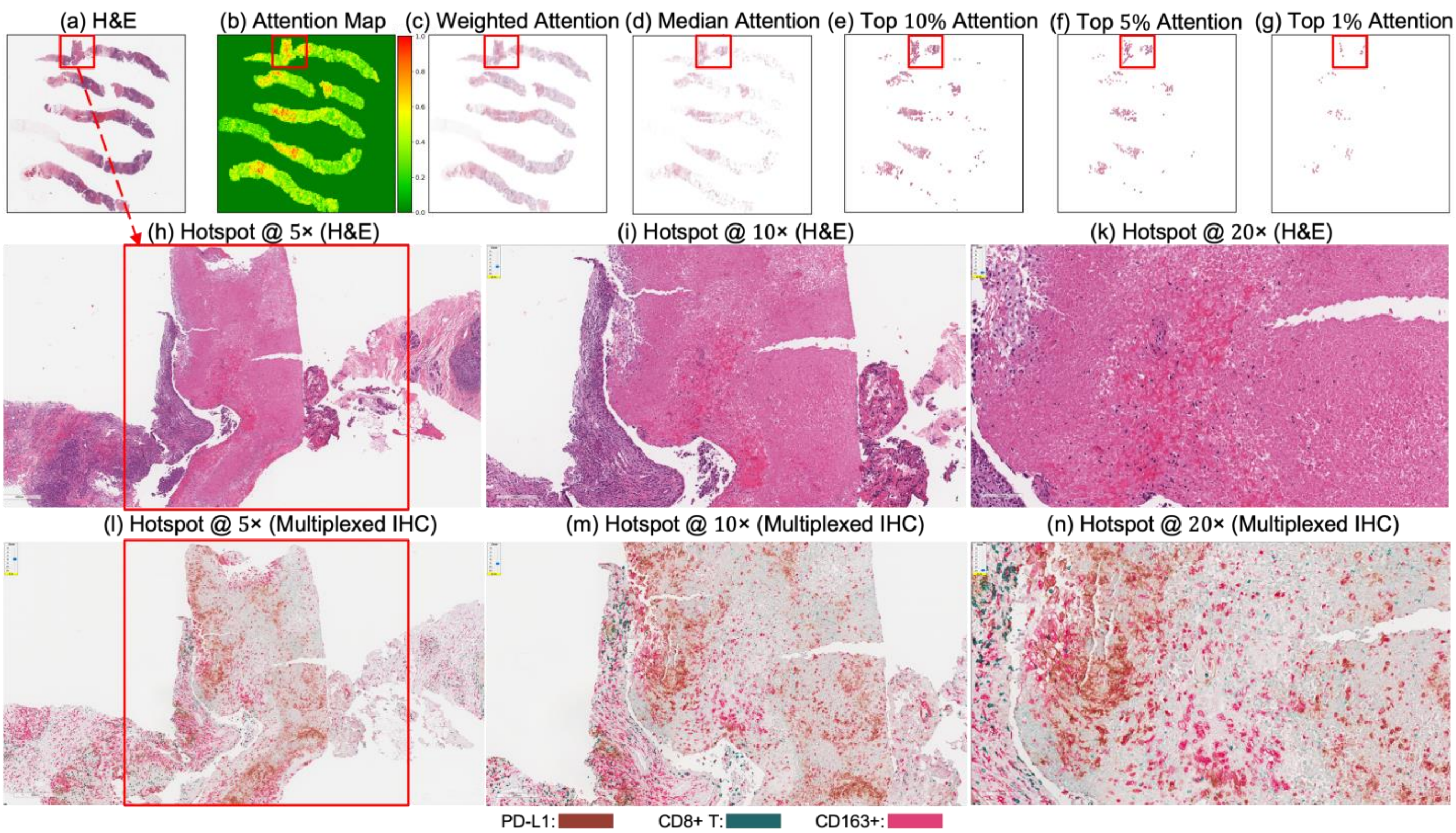

**Figure 4.** Attention map visualization of an attention-based multiple instance learning (MIL) model for a correctly classified triple-negative breast cancer (TNBC) patient (true positive) who achieved pCR to neoadjuvant chemotherapy (NACT). (**a**) H&E-stained biopsy slide thumbnail with (**b**) corresponding attention heatmap of the MIL model. (c) Weighted attention heatmap representation

showing individual patches weighted by the model's attention scores. (**d**) Median attention heatmap. (**e**–**g**) Progressive filtering of attention regions showing median attention: (**e**) top 10% attention (**f**) and top 5% attention, (**g**) culminating in top 1% attention hotspots. (**h**–**j**) Zoomed-in H&E slides of the identified hotspot region at increasing magnifications: 5×, 10×, and 20×, respectively. (**k**–**m**) Multiplex immunohistochemistry (mIHC) slides of consecutive tissue sections from the same hotspot region at the same magnifications (5×, 10×, 20×), revealing the presence of PD-L1 (brown), CD8+ T cells, and CD163+ macrophages (red) in the model-identified regions. These immune markers are established biomarkers for pCR in TNBC [27], demonstrating the model's ability to attend to immunologically relevant regions rich in biomarkers.

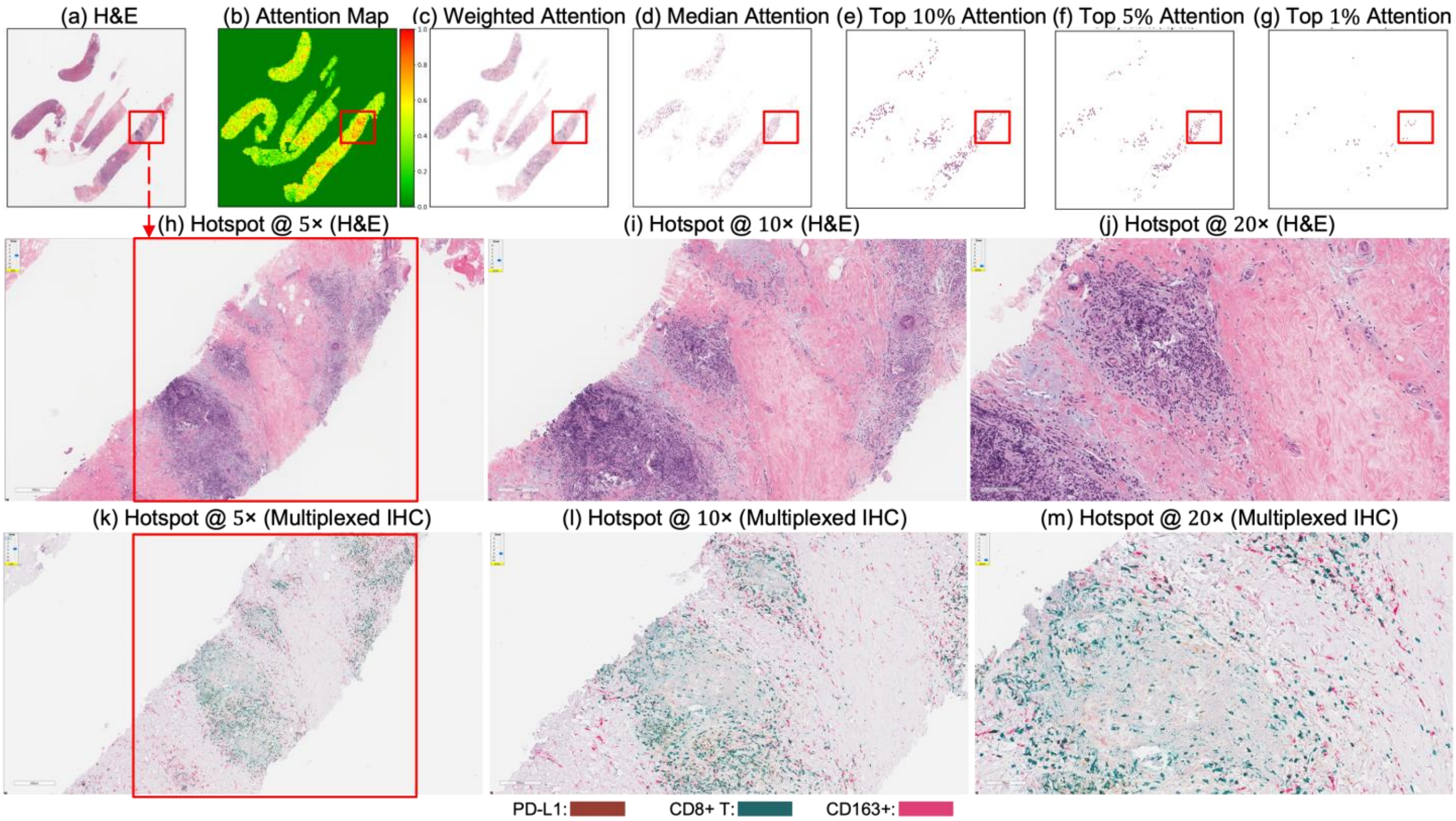

**Figure 5.** Attention map visualization of an attention-based multiple instance learning (MIL) model for a correctly classified triple-negative breast cancer (TNBC) patient (true negative) who did not achieve pathological complete response (non-pCR) to neoadjuvant chemotherapy (NACT). (**a**) H&E-stained biopsy slide thumbnail with (**b**) corresponding attention heatmap generated by the MIL model. (**c**) Weighted attention representation showing individual patches weighted by the model's attention scores. (**d**) Median attention. (**e**–**g**) Progressive filtering of attention regions showing median attention: (**e**) top 10% attention (**f**) and top 5% attention, (**g**) culminating in top 1% attention hotspots. (**h**–**j**) Zoomed-in H&E slides of the identified hotspot region at increasing magnifications: 5×, 10×, and 20×, respectively. (**k**–**m**) Multiplex immunohistochemistry (mIHC) slide of consecutive tissue sections from the same hotspot region at the same magnifications (5×, 10×, 20×), revealing the presence of PD-L1 (brown), CD8+ T cells, and CD163+ macrophages (red) in the model-identified regions. These immune markers are established biomarkers for pCR in TNBC [27], demonstrating the model's ability to attend to immunologically relevant regions rich in biomarkers.

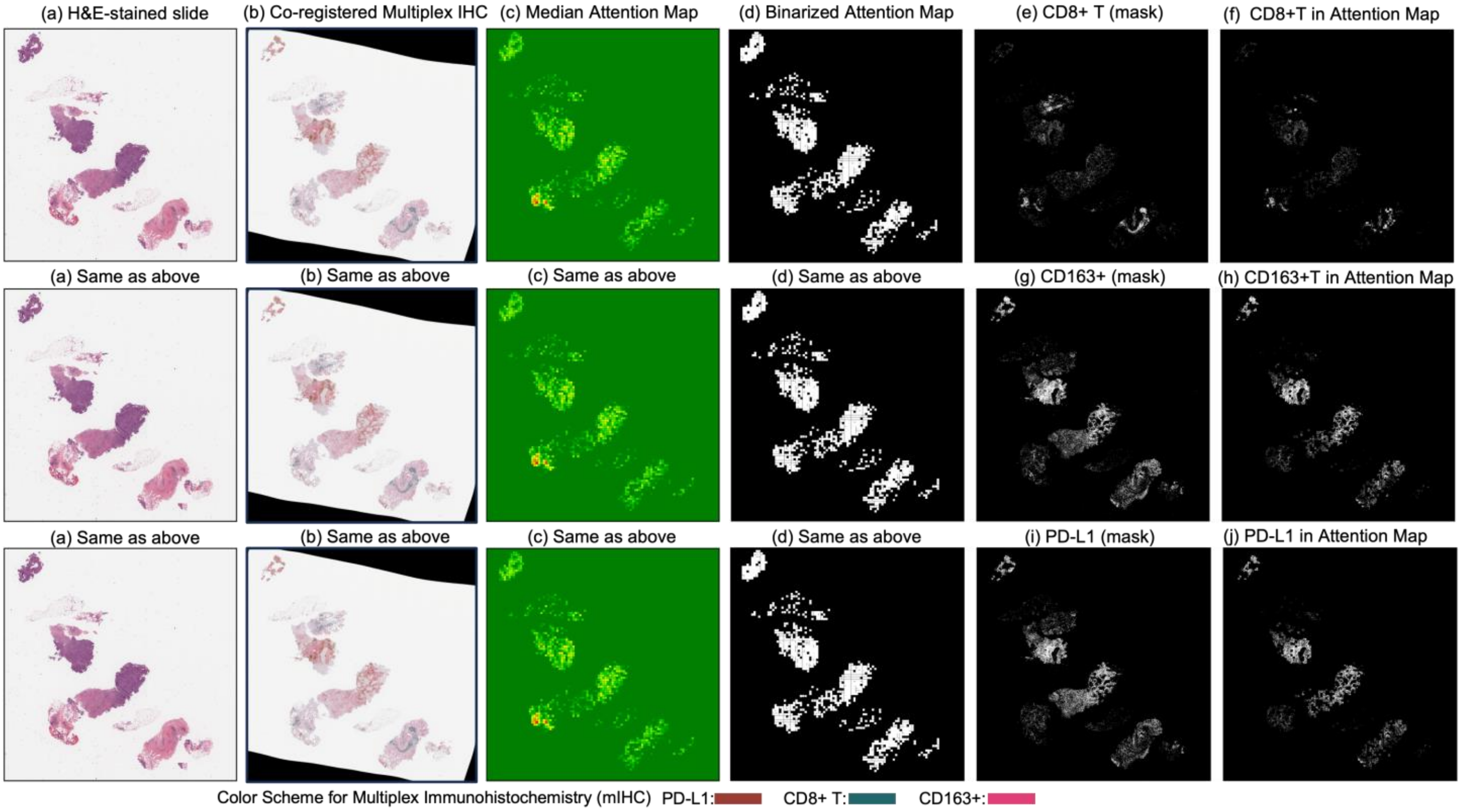

**Figure 6.** Columns (**a**) through (**d**) display (**a**) the original H&E-stained biopsy slide, (**b**) the corresponding co-registered multiplex immunohistochemistry (mIHC) slide, (**c**) the median attention map generated by the attention model, and (**d**) the binarized version of the attention map. Column (**e**) shows the CD8+ T-cell mask, and column (**f**) illustrates the intersection between the binarized attention map (**d**) and the CD8+ T-cell mask (**e**), indicating the presence of CD8+ T cells within the model's attention regions. Similarly, column (**g**) presents the CD163+ cell mask, and column (**h**) shows the intersection between (**d**) and (**g**), reflecting the attention overlap with CD163+ regions. Column (**i**) displays the PD-L1 mask, and column (**j**) presents the intersection between (**d**) and (**i**), quantifying the presence of PD-L1 in the attended regions.

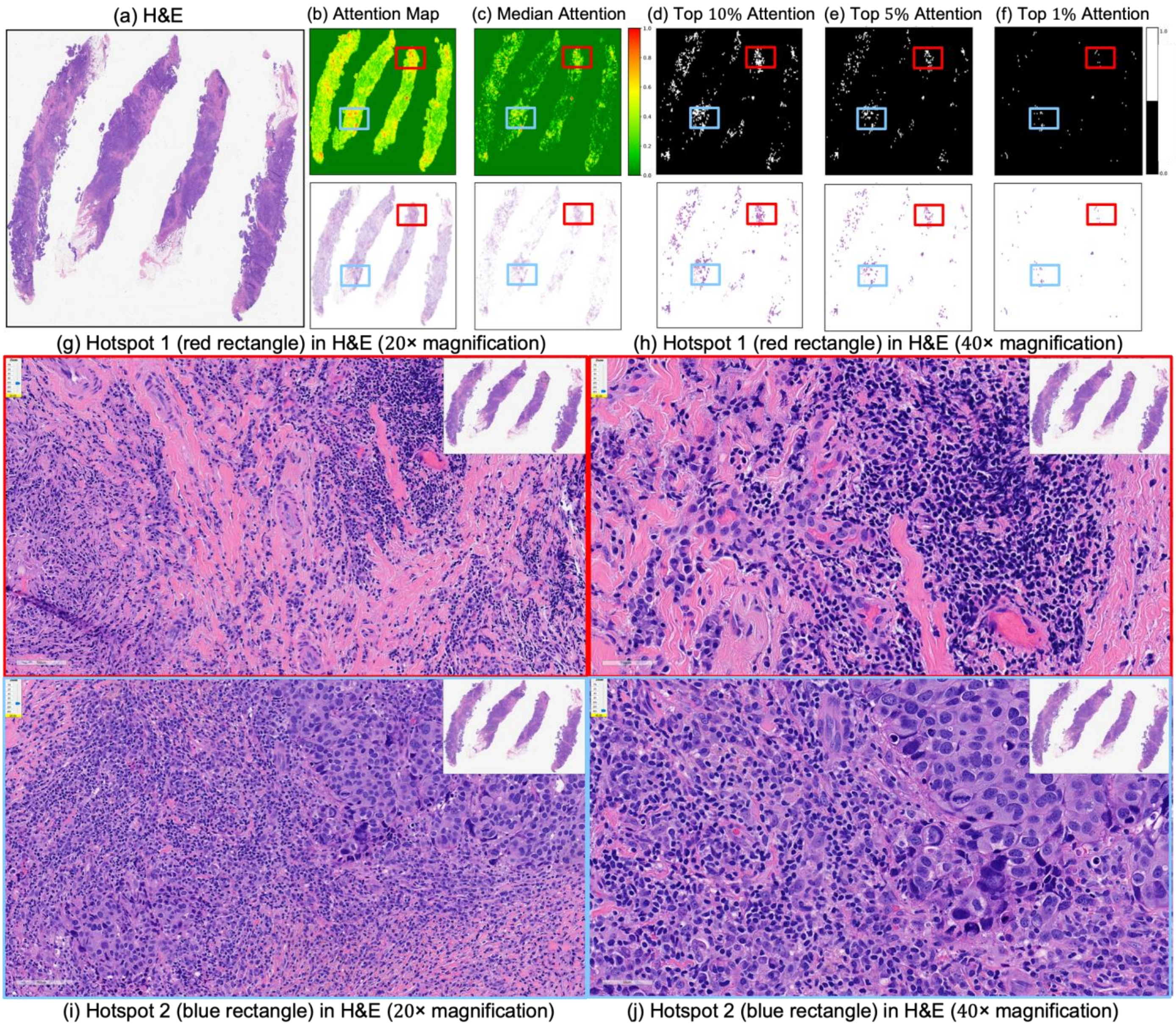

**Figure 7.** Attention map visualization for an incorrectly classified triple-negative breast cancer (TNBC) patient who achieved pathological complete response (pCR) to neoadjuvant chemotherapy (NACT) but who the model predicted as non-pCR. (**a**) H&E-stained biopsy slide thumbnail. (**b**,**c**) First row: corresponding attention heatmap generated by the deep learning model. The second row displays the weighted attention representation showing individual patches weighted by the model's attention scores. (**c**) Median attention. (**d**–**f**) display the top 10%, 5%, and 1%, attention masks while below each mask is shown individual patches weighted by the model's attention scores. (**g**,**h**) show the zoomed-in H&E slide of the identified hotspot region 1 (highlighted by the red rectangle) at increasing magnifications of 20×, and 40×, respectively. (**i**,**j**) show the zoomed-in H&E slides of the identified hotspot region 2 (highlighted by the blue rectangle) at increasing magnifications of 20×, and 40×, respectively.

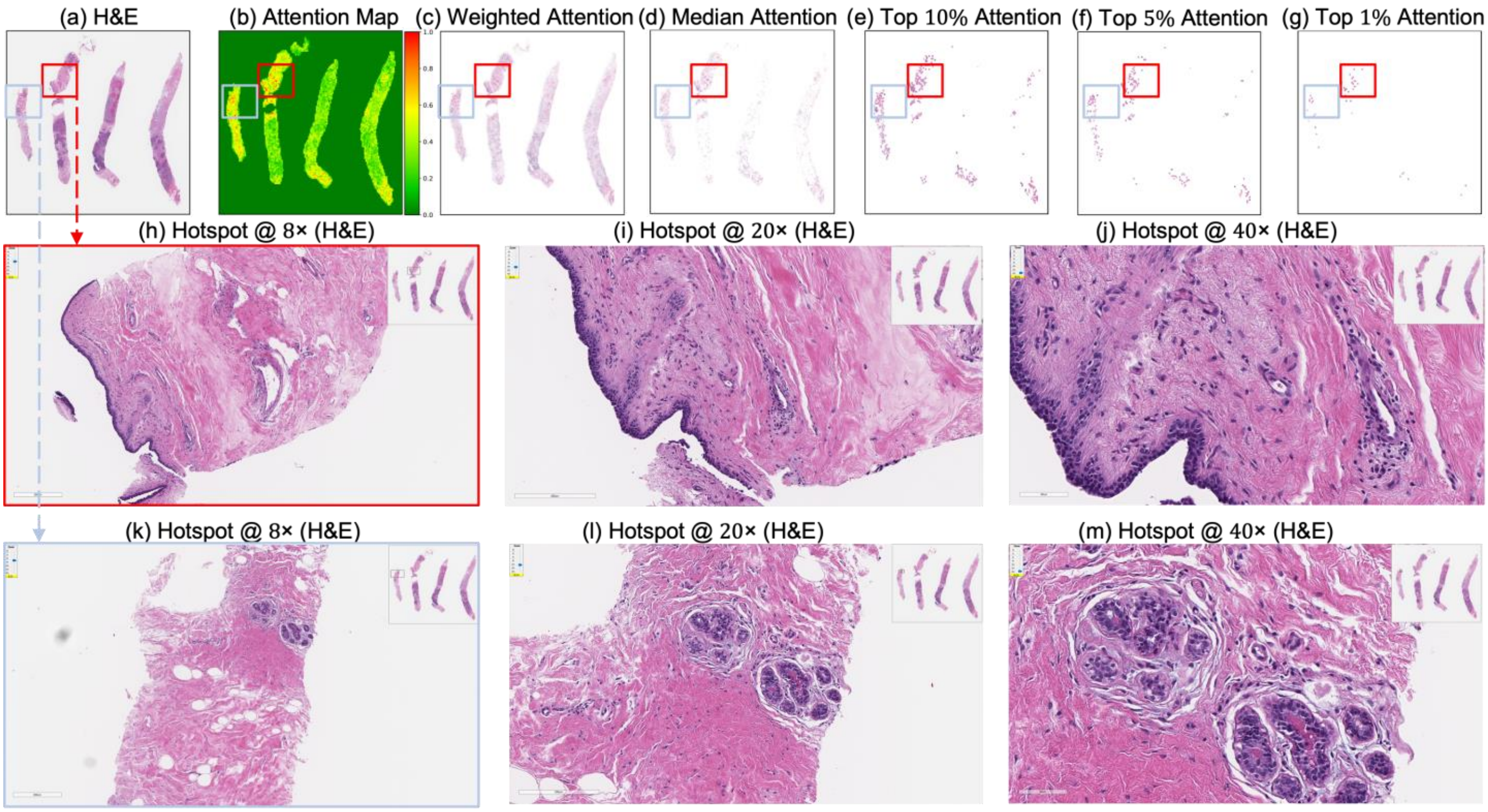

**Figure 8.** Attention map visualization of an attention-based multiple instance learning (MIL) model for an incorrectly classified TNBC patient (false negative) who achieved pathological complete response (pCR) to neoadjuvant chemotherapy (NACT), but who the model predicted as non-pCR. (**a**) H&E-stained biopsy slide thumbnail with (**b**) corresponding attention heatmap of the MIL model. (**c**) Weighted attention representation showing individual patches weighted by the model's attention scores. (**d**) Median attention. (**e**–**g**) Progressive filtering of attention regions showing median attention, (**e**) top 10% attention, (**f**) and top 5% attention, (**g**) culminating in top 1% attention hotspots. (**h**–**k**) Zoomed-in H&E slide of the identified hotspot region (highlighted by the red rectangle) at increasing magnifications: 8×, 20×, and 40×, respectively. (**l**–**n**) Zoomed-in H&E slides of the identified hotspot region (highlighted by the sky-blue rectangle) at increasing magnifications: 8×, 20×, and 40×, respectively.

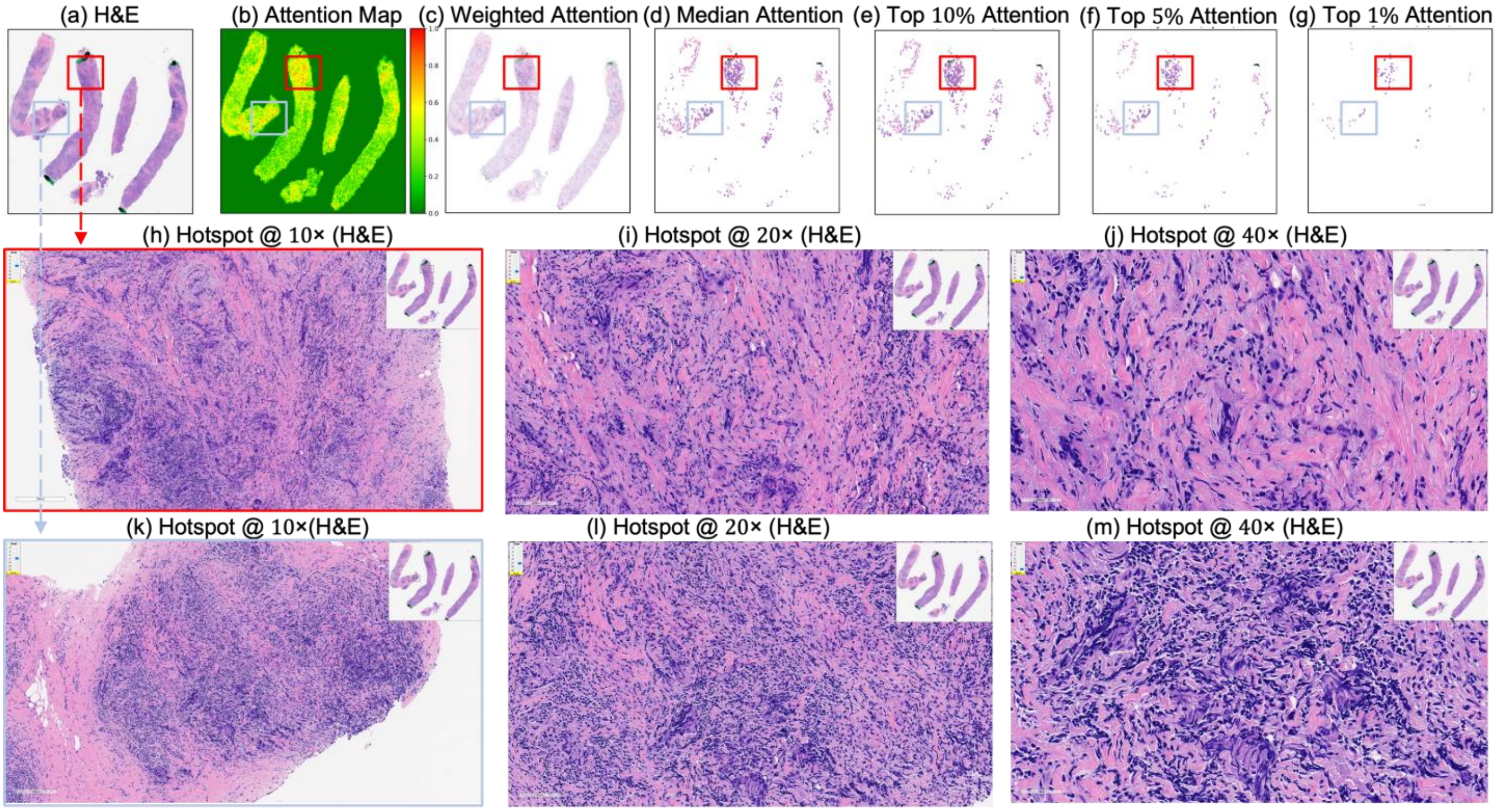

**Figure 9.** Attention map visualization of an attention-based multiple instance learning (MIL) model for an incorrectly classified triple-negative breast cancer (TNBC) patient (false positive) who did not achieve pathological complete response (non-pCR) to neoadjuvant chemotherapy (NACT), but who the model predicted as pCR. (**a**) H&E-stained biopsy slide thumbnail with (**b**) corresponding attention heatmap of the MIL model. (**c**) Weighted attention representation showing individual patches weighted by the model's attention scores. (**d**) Median attention. (**e**–**g**) Progressive filtering of attention regions showing median attention, (**e**) top 10% attention, (**f**) and top 5% attention, (**g**) culminating in top 1% attention hotspots. (**h**–**j**) Zoomed-in H&E slide of the identified hotspot region (highlighted by the red rectangle) at increasing magnifications: 10×, 20×, and 40×, respectively. (**k**–**m**) Zoomed-in H&E slide of the identified hotspot region (highlighted by the sky-blue rectangle) at increasing magnifications: 10×, 20×, and 40×, respectively.

### *6.5. Significance and Clinical Implications*

The ability to predict NACT response in TNBC holds significant clinical value, enabling more personalized and timely treatment decisions [2-6]. Approximately 40–50% of TNBC patients achieve pCR, which correlates strongly with improved survival outcomes. In contrast, those with residual disease (non-pCR) have a higher risk of recurrence and mortality, highlighting the importance of early pCR prediction to NACT. Our framework demonstrates robust predictive performance and biological interpretability and is designed for seamless integration into existing digital pathology workflows. At The Ohio State University Wexner Medical Center, one of the largest academic cancer centers in the U.S. and a recognized leader in digital pathology, our model can be embedded into whole-slide imaging systems, providing pathologists with interpretable predictions to support treatment planning. Ultimately, our work advances the integration of AI into clinical practice, aiding oncologists and pathologists in making timely, patient-specific decisions, especially for aggressive cancers like TNBC.

## 7. Conclusions

In this study, we present an attention-based MIL framework to predict treatment response (i.e., pCR) to NACT in patients with TNBC using routine pre-treatment H&E-stained biopsy slides. The framework demonstrated strong predictive performance across both internal and external cohorts, with high generalizability despite the inherent heterogeneity of TNBC. Notably, the integration of attention mechanisms and the availability of mIHC slides for a subset of patients enabled spatial interpretability, revealing alignment between high-attention regions and immune biomarkers, PD-L1, CD8+ T cells, and CD163+ macrophages, validated through aligned mIHC. The moderate overlap (mean IoU ≈ 0.46) between attention maps and immune-enriched regions underscores the biological relevance of the model's predictions. These findings highlight the potential of our framework to support personalized treatment planning, reduce overtreatment, and accelerate biomarker discovery in TNBC. Future work will focus on validating this approach in larger, multi-institutional cohorts, integrating clinical and genomic variables, and extending its application to other breast cancer subtypes and treatment response endpoints.

**Supplementary Materials:** The following supporting information can be downloaded at https://www.mdpi.com/article/doi/s1: Supplementary Table 1 details the distribution of patient cohorts used in the study, including the number of pathological complete and incomplete response cases for both the in-house OSU Wexner Medical Center cohort and the independent MD Anderson Cancer Center cohort. Supplementary Table 2 lists the software libraries and packages with versions used to ensure reproducibility. Supplementary Tables 3 and 5 provide the grid search ranges for hyperparameters explored during model selection for both the attention-based MIL and traditional ML models, respectively. Supplementary Tables 4 and 6 present the final selected hyperparameters for those models..

**Author Contributions:** H.K. led the work, designed the experiments, analyzed the data, prepared the figures and tables, and wrote the manuscript. Z.L. and M.K.K.N. obtained the study cohort. Z.S., H.Z., Y.W., B.N., S.W., and H.G. reviewed the manuscript and provided valuable insights and comments. All authors have read and agreed to the published version of the manuscript.

**Funding:** This project was supported by R01 CA276301 (PIs: Niazi, Chen) from the National Cancer Institute. The project was also supported by The Ohio State University Comprehensive Cancer Center, Pelotonia Research Funds and the Department of Pathology. The content is solely the responsibility of the authors and does not necessarily represent the official views of the National Institutes of Health or National Cancer Institute.

**Institutional Review Board Statement:** This study was approved by the Institutional Review Board (IRB Protocol #2016C0025).

**Informed Consent Statement:** All procedures performed in studies involving human participants were in accordance with the ethical standards of the institutional and/or national research committee and with the 1964 Helsinki declaration and its later amendments or comparable ethical standards. Informed consent was obtained from all individual patients included in the study.

**Data Availability Statement:** Original data used in this study can be requested by emailing to the corresponding authors Hikmat Khan at hikmat.Khan@osumc.edu or Zaibo Li at zaibo.li@osumc.edu.

**Conflicts of Interest:** All authors have no conflicts of interest.

## Abbreviations

The following abbreviations are used in this manuscript:

| | |
|---|---|
| TNBC | Triple-negative breast cancer |
| H&E | Hematoxylin and eosin |
| AI | Artificial intelligence |
| CD8+ T | Cluster of differentiation 8-positive T cells |
| CD163+ | Cluster of differentiation 163-positive macrophages |
| PD-L1 | Programmed death-ligand 1 |
| ER | Estrogen receptor |
| PR | Progesterone receptor |
| HER2 | Human epidermal growth factor receptor 2 |
| NACT | Neoadjuvant chemotherapy |
| pCR | Pathological complete response |
| AUC | Area under the curve |
| IHC | Immunohistochemistry |
| TME | Tumor microenvironment |
| TILs | Tumor-infiltrating lymphocytes |
| OSU | The Ohio State University |
| MIL | Multiple instance learning |
| WSI | Whole-slide image |
| mIHC | Multiplex immunohistochemistry |
| TAMs | Tumor-associated macrophages |
| CAD | Computer-aided diagnosis |
| GPU | Graphics processing unit |
| VRAM | Video random-access memory |
| ROC | Receiver operating characteristic |
| UNI v2 | Self-supervised pathology foundation model |